# Distributed Acoustic Sensing Links Millisecond Lightning-Pulse Charge Transfer to Thunder-Induced Ground Motion

Heting Hong[1*], Yashun Tian[2], Gaopeng Lu[2*], Gan Yang[3], Shaodong Chen[3], Yanfeng Fan[4], Weitao Lyu[4], Jingxiao Liu[1*]

1. School of Civil and Environmental Engineering, Nanyang Technological University, Singapore 639798, Singapore.

2. School of Earth and Space Sciences, University of Science and Technology of China, Hefei 230026, China.

3. Institute of Tropical and Marine Meteorology, China Meteorological Administration, Guangzhou 510640, China,

4. State Key Laboratory of Severe Weather Meteorological Science and Technology / CMA Key Laboratory of Lightning, Chinese Academy of Meteorological Sciences, Beijing, China

**Corresponding authors:**

Dr. Heting Hong (heting.hong@ntu.edu.sg; hardyhong1314@gmail.com);

Prof. Gaopeng Lu (gaopenglu@gmail.com);

Prof. Jingxiao Liu (jingxiao.liu@ntu.edu.sg).

## Abstract

Overlapping arrivals from extended lightning channels obscure links between individual electrical processes and ground motion. We combine distributed acoustic sensing (DAS) with synchronized current, optical, nodal-seismometer, and infrasound measurements of rocket-triggered lightning. DAS locates 47 low-altitude acoustic sources across seven flashes, including sources associated with initial current variation, individual initial continuous current pulses, and return strokes. The horizontal bootstrap stability radius is 2.7 m under fixed model assumptions. For 35 sources from five flashes with synchronized current records, first-millisecond pulse charge transfer correlates positively with distance-corrected DAS and nodal vertical-velocity measures. Site-specific transfer functions calibrated on three flashes reconstruct principal pressure oscillations in a held-out flash from vertical ground velocity in the 28–34 Hz band. These observations link early charge transfer to event-specific ground responses and show that thunder-induced ground motion retains information about relative electrical variations within flashes.

## Plain Language Summary

Lightning rapidly heats the air, producing thunder that can also shake the ground. However, overlapping sounds from different parts of a lightning channel make it difficult to connect individual electrical pulses with ground motion. We combined measurements along a fiber-optic cable with electrical current, high-speed images, ground vibration, and air pressure during seven rocket-triggered lightning flashes. The cable measurements located 47 low-altitude thunder sources. Among events within the same flash, greater charge transfer during the first millisecond was associated with stronger ground responses. At two sites, relationships calibrated using three flashes reproduced the main pressure oscillations of a different flash from its ground vibration within a limited frequency band. These findings show how coordinated measurements can reveal electrical variations within lightning flashes and help interpret the ground shaking caused by thunder.

## Key Points

- Dense DAS locates 47 low-altitude acoustic source events from seven triggered-lightning flashes with meter-scale uncertainty.
- First-millisecond charge transfer, $Q_{1\mathrm{ms}}$, correlates with distance-corrected DAS and vertical nodal-seismometer (Node-Z) responses.
- Colocated pressure and vertical ground-velocity measurements reveal site-specific thunder acoustic–ground coupling.

## Keywords

Distributed Acoustic Sensing (DAS); thunder; triggered lightning; acoustic-to-seismic coupling; source location; charge transfer.

## 1 Introduction

Thunder originates from the rapid heating and expansion of a geometrically complex lightning channel. Studies of acoustic spectra, channel reconstructions, and return-stroke models show that a received waveform reflects source energetics, propagation effects, and the superposition of distributed channel segments (Few, 1969; Few, 1970; Holmes et al., 1971; Plooster, 1971; Ribner & Roy, 1982). Close-range measurements and subsequent modeling further show that thunder characteristics depend on electrical parameters and the evolution of the discharge along the lightning channel, while waveform simulations and sound-synthesis studies examine how thunder characteristics can be reproduced (Bocharov et al., 2024; Depasse, 1994; Farges & Blanc, 2010; Petculescu & Kruse, 2014; Reiss et al., 2021). A single electrical process can generate acoustic radiation from multiple segments of an extended lightning channel. Signals from segments at similar source–receiver distances may arrive nearly simultaneously, while electrical processes occurring in rapid succession within the same flash further compound this overlap. Consequently, individual waveform features in a single-receiver record can be difficult to associate with specific electrical processes.

Acoustic arrays can constrain source location and timing through differences in arrival times across receivers. Triggered lightning provides a well-constrained setting in which acoustic source estimates can be evaluated against independently recorded electrical and optical observations (Lu et al., 2014; Zhang et al., 2016). The early current record of a negative rocket-triggered lightning flash can include an initial current variation (ICV), associated with current changes as the triggering wire disintegrates and is replaced by a plasma channel (Rakov et al., 2003; D. Wang et al., 1999). The initial continuous current (ICC) may contain superimposed ICC pulses (ICCPs). Subsequent return strokes (RSs) may occur after the ICC ceases, while M-components can occur during continuing-current stages (Qie et al., 2014). Previous DAS observations have identified thunder associated with continuing-current activity and M-components (H. Hong et al., 2025), but resolving distinct acoustic sources associated with ICV and individual ICCP events remained unclear.

In such triggered lightning experiments, thunder has been located using microphones and joint acoustic–electromagnetic observations (Anderson et al., 2014; Arechiga et al., 2011), while infrasound and acoustic imaging have identified spatially distinct channel activity (Arechiga et al., 2014; Dayeh et al., 2015). Lightning mapping and acoustic arrays have also enabled statistical reconstruction of storm discharges (Gallin et al., 2016). These approaches have supported estimates of acoustic spectra and energy (Lacroix et al., 2018; Lacroix et al., 2019), as well as the vertical distribution of sound power (Bestard et al., 2023; Bestard et al., 2025). However, source overlap and receiver geometry continue to limit thunder-based channel reconstruction (J. Wang et al., 2022), particularly when closely spaced electrical processes produce acoustic arrivals that cannot be separated. Relating resolvable thunder events to electrical activity therefore requires both spatial constraints on the acoustic sources and synchronized measurements of the associated electrical processes.

Thunder pressure can also load the ground and generate ground motion. Such signals are documented in conventional seismic records (Kappus & Vernon, 1991; Lin & Langston, 2007; Lin & Langston, 2009a; Lin & Langston, 2009b), and acoustic-to-seismic coupling depends on frequency, shallow structure, and receiver installation (Anthony et al., 2022; Novoselov et al., 2020). Regional seismoacoustic observations and analyses of thunder-induced seismic spectra further document this coupling at larger scales (T.-K. Hong et al., 2023; Novoselov et al., 2022). DAS adds dense spatial sampling and has mapped thunder-induced ground motion and reconstructed three-dimensional thunder-source locations (H. Hong et al., 2024; H. Hong et al., 2025; Zhu & Stensrud, 2019). Building on these advances, a remaining challenge is to quantify how ground responses associated with resolvable thunder events vary with lightning electrical activity, while independently evaluating local pressure–ground coupling. These complementary observations are helpful for distinguishing source-related variability from acoustic propagation and local ground-coupling effects, and thereby can assess how reliably thunder-induced ground motion reflects lightning electrical activity at the source scale.

To examine this relationship, we combine synchronized DAS, current, optical, nodal-seismometer, and infrasound observations of triggered lightning. We use DAS to locate low-altitude acoustic sources, each representing a resolvable thunder event that may include closely spaced electrical processes whose acoustic contributions cannot be separated. For each matched event, we characterize early electrical activity using $Q_{1\text{ms}}$, the charge transferred at the channel base over a fixed 1-ms window. This measure includes current contributions beyond the instantaneous peak and has precedent in triggered-lightning return-stroke studies (Schoene et al., 2010). It provides a common integration duration for comparing early charge transfer, but does not directly measure total event charge, channel heating, or radiated acoustic energy. Charge distribution along the channel and nonlinear channel resistance further complicate the relationships among channel-base current, remote electromagnetic fields, and radiated energy (da Silva et al., 2019; Thottappillil et al., 1997). We show that greater $Q_{1\text{ms}}$ is associated with stronger distance-corrected DAS and vertical nodal-seismometer (Node-Z) ground-motion responses after within-flash centering. At two colocated sites, pressure and vertical ground velocity covary in the dominant frequency band of thunder, and site-specific transfer functions reconstruct the principal pressure oscillations of a held-out flash from ground velocity. Together, these results connect early lightning charge transfer with event-specific ground responses and provide a basis for using thunder-induced ground motion to investigate relative electrical variations among resolvable thunder events within individual flashes, provided that source geometry and local pressure–ground coupling are constrained.

## 2 Data and methods

### 2.1 Triggered lightning observations

The field experiment was conducted in 2024 at the Guangzhou Experiment Site for Triggered Lightning and Testing. The observation system combined a 506-channel DAS cable sampled at 5 kHz with 2 m channel spacing, 250-Hz nodal seismometers, a 70,000-fps high-speed camera, a 50-kA-range channel-base current meter sampled at 10 MHz, and two 100-Hz infrasound sensors (Figure 1). The July and August Node deployments contained 24 and 23 stations, respectively. The camera was located approximately 1.9 km southwest of the experimental site; the channel-base current meter was at DAS Ch505; and infrasound sites 203 and 204 were at DAS Ch257/Node 6053 and DAS Ch424/Node 6164, respectively. Across seven triggered-lightning flashes, we retained 47 low-altitude thunder acoustic source events with clear and separable DAS signals suitable for location: 3, 14, 10, 6, 3, 7, and 4 sources in Flashes 1–7 (F1–F7), respectively. All source identifiers in the paper follow this Flash 1–7 catalog (Table S1).

Figure 2 shows representative multisensor signatures associated with the retained sources in Flash 4. In the 40–450 Hz DAS strain-rate wavefield, F4-S01–F4-S06 form distinct, spatially coherent acoustic moveouts across the cable rather than isolated responses at individual channels, providing the separability required for source location (Figure 2a). The corresponding channel-base current record contains a sequence of impulsive current pulses with different amplitudes, accompanied by discrete optical brightening (Figure 2b). After acoustic propagation, pressure fluctuations occur near the geometry-predicted arrivals at the infrasound sites, while colocated DAS and Node-Z sensors record temporally consistent ground motion (Figure 2c). These observations provide a common timing framework linking individual electrical and optical features to the acoustic wavefield and its ground response. Sensor availability and analysis-valid quantities for all 47 sources are summarized in Table S1; unavailable or quantitatively invalid measurements are treated as missing rather than assigned zero values.

## 2.2 Thunder source location and sensitivity analysis

DAS records were median-centered, corrected for common-mode noise, band-pass filtered between 40 and 450 Hz to suppress low-frequency anthropogenic and high-frequency background noise, and converted to robustly normalized Hilbert-envelope images. Channels 0–25, 149–238, and >490 showed poor cable–ground coupling and were excluded, leaving 375 usable channels (Figure 2a). For a source at position $r_s$ and origin time $t_0$, the predicted acoustic arrival at DAS channel $i$ is

$$\hat{t}_i = t_0 + \frac{\|r_i - r_s\|}{c}, \quad (1)$$

where $r_i$ is the three-dimensional channel position and $c = 343\ \mathrm{m\,s^{-1}}$. Source location was based on DAS backprojection (Li et al., 2023), with additional high-signal-to-noise timing constraints introduced only for sources with less distinct DAS signatures. Candidate locations were evaluated using the normalized Hilbert-envelope response along the predicted arrival curve across the usable DAS channels (see Text S1 for the full location procedure; workflow summarized in Figure S1). The DAS wavefields for all seven flashes, including the coherent moveouts used for location, are shown in Figure S2.

**2.3 Charge transfer and ground response metrics**

For each current record with a measurable channel-base current and no clipping, $Q_{1ms}$ was calculated by integrating the baseline-corrected unidirectional current over a fixed 1-ms interval following the current-pulse onset. This metric provides a common-duration measure of early charge transfer that includes current contributions beyond the instantaneous peak, but does not represent the total event charge or acoustic energy (Schoene et al., 2010). In contrast, because the relationship between current and optical emission varies among discharge processes (Zhou et al., 2013), optical brightness was relatively calibrated, reported in relative units, and used only for timing rather than as a quantitative measure of electrical activity (Figure S3).

Ground response was quantified independently using DAS and Node-Z records (Figure 3). The DAS metric was defined as the receiver median of $r\sqrt{\int \dot{\varepsilon}\,(t)^2\,dt}$ from 5–500 Hz strain-rate data, whereas the Node-Z metric was the station median of $r\max(|\,v_z\,|)$ from response-removed 5–120 Hz vertical velocity. Here, $r$ is the three-dimensional source–receiver distance and provides a first-order $1/r$ geometric-spreading correction. The DAS metric was chosen to robustly characterize the full oscillatory strain-rate wave packet, whereas peak vertical velocity provides a direct measure of the strongest ground motion recorded by the conventional seismic sensors. The frequency bands were selected to retain the principal thunder-related ground response while suppressing low-frequency background motion and frequencies affected by instrumental resonances or noise. Medians across receivers or stations were used to reduce sensitivity to local coupling variations and anomalous records.

### 2.4 Statistical analysis of charge–response relationships

We evaluated the relationships between first-millisecond charge transfer and the distance-corrected DAS and Node-Z responses for 35 source events from Flashes 1–5. For within-flash inference, flash-specific means were removed from the $\log_{10}$-transformed charge and response values. Slope uncertainty was estimated from 20,000 source bootstrap resamples stratified by flash, with 95% confidence intervals obtained from the bootstrap distribution. Spearman correlation quantified the monotonic association, and significance was assessed using 50,000 within-flash permutations of response values with a two-sided test.

## 3 Results

### 3.1 DAS locates 47 low-altitude thunder sources

The 506-channel DAS array located 47 low-altitude thunder sources across seven flashes (Figure 1; Table S1; F1-F7 in Figure S2). We retained only sources with clear, separable coherent DAS moveouts and sufficient temporal distinction to be treated as individual thunder-producing processes. The located sources are concentrated in the vicinity of the lightning channel. To quantify the sensitivity of the source locations to DAS-channel sampling, the retained channels were divided into 10 contiguous blocks, and 10 blocks were sampled with replacement in each of 100 replicates per source, while source time, sound speed, cable geometry, and processing settings were held fixed. The resulting 4,700 relocations yielded a 95th-percentile horizontal offset of 2.7 m relative to the corresponding full-channel solutions. Previous triggered-lightning experiments at this site also demonstrated meter-scale near-ground DAS location accuracy against known rocket and wire positions (H. Hong et al., 2025). In addition to the strongest acoustic arrivals associated with RSs, distinct DAS-resolved acoustic arrivals are individually matched to ICVs and individual ICC pulses (Figures S2 and S3). These observations provide direct event-level correspondence between ICVs or individual ICC pulses and distinct acoustic sources. The ICV is primarily associated with current changes in the triggering wire. M-components are also identified in the electrical records, but their acoustic contributions cannot be independently resolved due to temporal overlap with those of nearby RSs (Figure S3).

Figure 2 illustrates the multisensor signatures of the source events in Flash 4. These events correspond to distinct, coherent DAS moveouts and identifiable features in the current and optical records. At the colocated sites, infrasound and ground-motion signals occur near the geometry-predicted acoustic arrivals without additional fitted time shifts.

Having established the DAS source catalog and assessed its location stability, we can examine the relationships among electrical, optical, acoustic, and ground-motion signatures associated with individual resolvable thunder events. The consistent timing across sensors provides the basis for the event-level charge–response comparisons in the following sections.

### 3.2 First-millisecond charge transfer covaries with ground response

We analyzed 35 source events from Flashes 1–5 with valid, unsaturated current records and the ground-response quantities used in Figure 3 (Table S1). For these events, first-millisecond charge transfer $Q_{1\mathrm{ms}}$ is positively associated with both distance-corrected DAS and Node-Z responses (Figures 3a and 3b). To evaluate whether these associations persist within flashes, we centered the $\log_{10}$-transformed charge and response variables by subtracting their respective flash-specific means. Both relationships remain positive after this within-flash centering (Figures 3c and 3d). The within-flash Spearman correlations between $Q_{1\mathrm{ms}}$ and the distance-corrected ground-response measures are 0.70 for DAS and 0.71 for Node-Z, with two-sided within-flash permutation $P$ values of $2.0 \times 10^{-5}$ and $4.0 \times 10^{-5}$, respectively. Regressions of the centered $\log_{10} Q_{1\mathrm{ms}}$ values on the corresponding centered $\log_{10}$ transformed ground responses yield slopes of 0.94 for DAS (95% stratified-bootstrap interval, 0.66–1.19) and 0.75 for Node-Z (0.45–1.06). These comparable correlations across two distinct ground-motion measures indicate that, within individual flashes, greater first-millisecond charge transfer is associated with stronger ground responses.

### 3.3 Narrow-band acoustic–ground coupling

For 16 temporally resolvable sources from Flashes 4–7, colocated infrasound and Node-Z measurements yielded 32 source–station pairs. We explored the pressure–ground relationship by comparing these records across several narrow frequency bands spanning 4–46 Hz in 6-Hz increments. Among the bands tested, 28–34 Hz exhibits the clearest common response and coincides with the dominant spectral energy of the N-wave-like thunder arrivals (Figure S4). We therefore use this band for the exploratory amplitude and waveform comparisons below. Within this band, Node-Z root-mean-square (RMS) velocity increases with infrasound RMS pressure (Figure 4a). The pooled empirical log–log fit has a slope of 0.674, with a residual standard deviation of 0.330 log units, equivalent to a multiplicative spread of approximately 2.14. This fit summarizes the amplitude covariation observed across the two sites in the selected band.

To further explore the relationship at the waveform level, we applied station-specific complex transfer functions estimated from Flashes 4, 5, and 7 to vertical ground velocity from the held-out Flash 6, without refitting. For F6-S02, the reconstructed equivalent pressure captures the principal oscillatory structure of the observed pressure at both sites (Figure 4b). Agreement is closer at site 203, where several major oscillations near and after the predicted arrival are reproduced in both phase and amplitude, whereas larger deviations occur at site 204 at later times. Differences in local acoustic–ground coupling and the complexity of the later wavefield may contribute to these discrepancies. The comparison preserves pressure amplitudes in pascals and uses the geometry-predicted arrival as the time reference, without RMS normalization or an additional fitted time shift. Despite the site-dependent differences, the agreement in waveform timing and structure indicates that the transfer functions can reproduce the main pressure oscillations of a flash not used for calibration. Equivalent comparisons for F6-S01–F6-S03 are shown in Figure S5.

## 4 Discussion

This study extends DAS-based thunder-source localization by linking early lightning charge transfer to event-specific ground responses and independently evaluating local pressure–ground coupling. DAS locates 47 low-altitude sources across seven flashes, providing arrival-time predictions and distance corrections that enable comparisons across electrical, optical, acoustic, and ground-motion records. Beyond the strong acoustic responses associated with RSs, we establish direct correspondence between distinct acoustic sources and ICVs and individual ICC pulses. These observations demonstrate that identifiable acoustic sources accompany multiple electrical processes in triggered-lightning experiments.

For 35 events from five flashes, greater first-millisecond charge transfer is associated with stronger distance-corrected DAS strain-rate and Node-Z vertical-velocity responses after within-flash centering. The comparable Spearman correlations of 0.70 and 0.71, respectively, strengthen the evidence that relative electrical variations are reflected in two distinct ground-motion measurements. At two colocated sites, pressure and vertical ground velocity covary in the 28–34 Hz band. Station-specific transfer functions calibrated on Flashes 4, 5, and 7 reproduce the principal pressure oscillations of held-out Flash 6 sources without refitting, demonstrating that locally calibrated ground velocity can recover key pressure-waveform features within this band. These complementary findings support the use of thunder-induced ground motion to investigate electrical variations among resolvable thunder events when source geometry and local pressure–ground coupling are constrained.

Several limitations arise from overlapping acoustic sources, fixed timing and propagation assumptions, and the limited coverage of events, sites, and frequencies. Several current peaks around F4-S02 occur within approximately 10 ms, but their individual acoustic contributions remain unresolved. The catalog therefore represents a resolvable subset of thunder-producing activity, and some events may contain multiple electrical processes. The 2.7 m bootstrap value describes horizontal location stability under channel resampling with fixed timing and propagation assumptions, rather than absolute accuracy. Moreover, $Q_{1\text{ms}}$ measures channel-base charge transfer rather than local heating or radiated acoustic energy.

Future work should address these limits through larger event samples, additional sites, and broader-band pressure measurements. Joint electrical, optical, and acoustic constraints could improve the separation of closely spaced events and clarify how their current histories relate to their acoustic signatures. Tests using alternative charge-integration windows and independent estimates of acoustic energy and propagation losses would help determine which electrical properties are most closely associated with ground response. Measurements across different ground conditions would also establish how pressure–ground transfer functions vary between sites and frequencies.

## 5 Conclusions

Although further work is needed to test these relationships across a wider range of events, sites, and frequencies, this study connects lightning physics with near-surface seismology through coordinated observations of electrical activity, acoustic pressure, and ground motion. By combining dense spatial sampling with direct measurements of atmospheric forcing, the approach brings source processes and local ground responses into a common observational framework. It supports interpreting thunder-induced ground motion as a complementary record of atmospheric activity, shaped by both the originating disturbance and its interaction with the ground. More broadly, this framework provides a basis for investigating rapid atmosphere–ground interactions through measurements across the interface.

## Acknowledgments

We thank Professor Baoshan Wang, Tao Fu, and Hongjie Zheng of the University of Science and Technology of China and Liwei Wang of the Guangdong Earthquake Agency for their help with data collection. This research was supported by the NTU Fellowships & Awards SUG for Research Award Number #025661-00022.

## Author contributions

H.H. conceived and designed the study, led and conducted the field experiments, performed the DAS source-location analysis, and wrote the manuscript. Y.T. analyzed the optical and current data. G.L. analyzed the optical and current data and contributed to manuscript writing. G.Y. and S.C. provided the current data. Y.F. and W.L. organized the field experiments and tests. J.L. analyzed the DAS data and contributed to manuscript writing. All authors reviewed the manuscript.

## Conflict of Interest

The authors declare there are no conflicts of interest for this manuscript.

## Open Research

### Data availability

The data supporting this study are available through Figshare (H. Hong, 2026) at https://doi.org/10.6084/m9.figshare.34020543.

### Software availability

DASView was used for selected DAS data processing, visualization, and figure preparation (H. Hong et al., 2026).

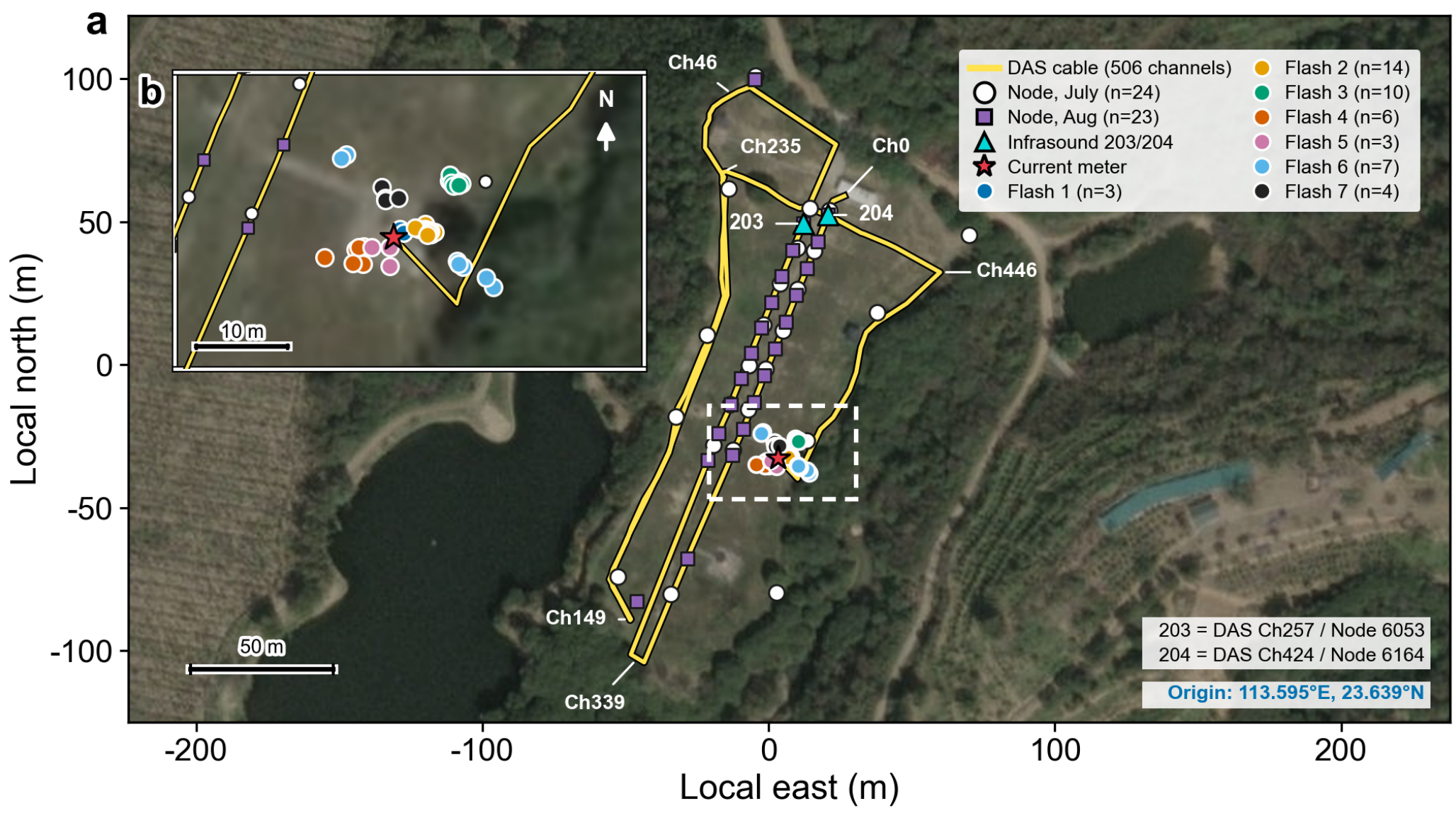

**Figure 1. Observation geometry and DAS-located low-altitude thunder sources.** (a) Site map showing the 506-channel DAS cable, July and August Node deployments, infrasound sites 203 and 204, the current meter at DAS Ch505, and 47 retained low-altitude thunder sources from seven flashes. Sites 203 and 204 are colocated with DAS Ch257 and Ch424 and Nodes 6053 and 6164, respectively. Colors denote Flash 1–7. (b) Enlarged view of the source cluster with unchanged source locations. Scale bars are 50 m and 10 m in (a) and (b), respectively.

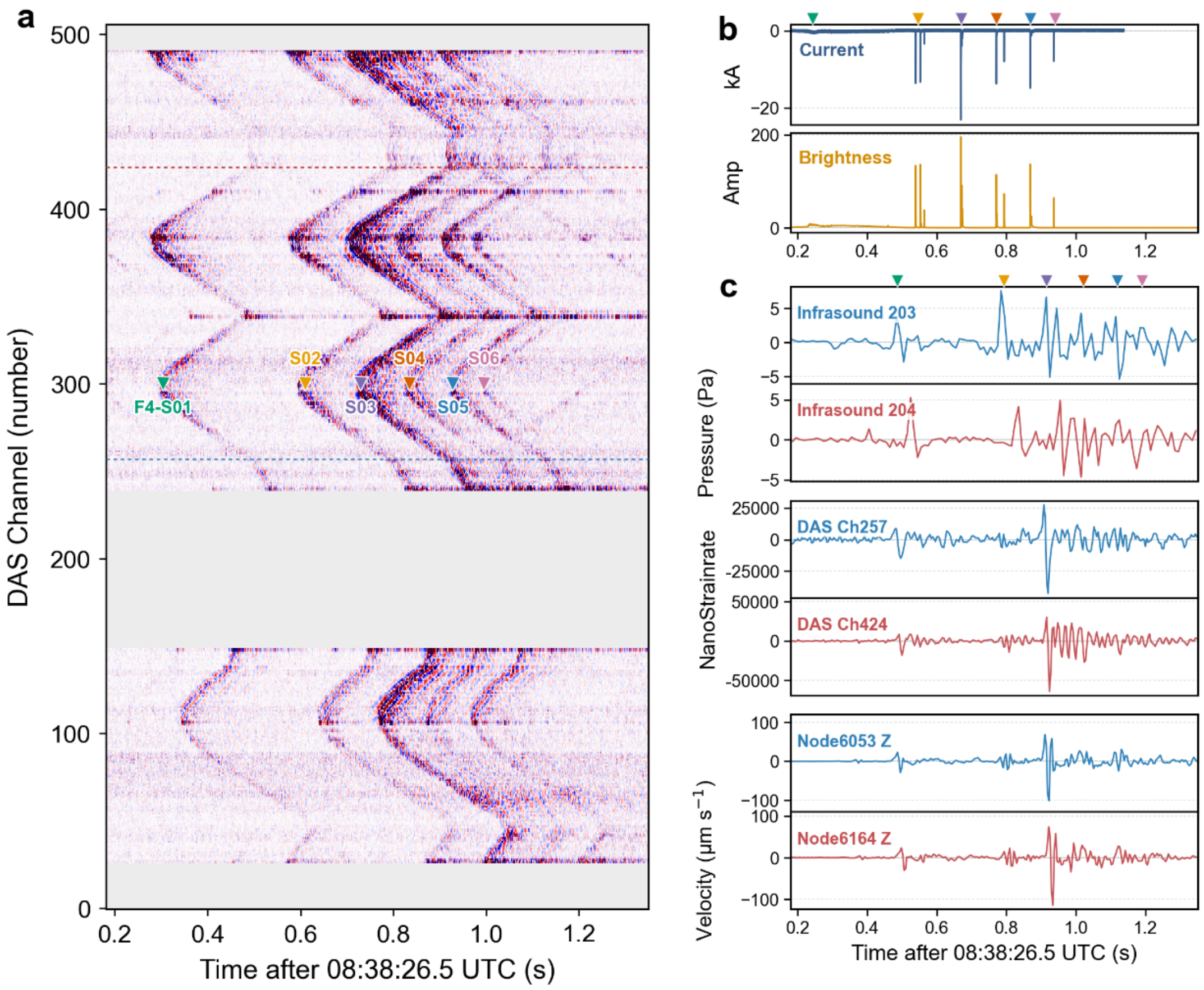

**Figure 2. Multisensor timing correspondence for six retained thunder source events in Flash 4.** (a) Signed DAS strain rate after common-mode removal and 40–450 Hz filtering. Light-gray gaps mark DAS channels not used in the analysis. Inverted triangles at Ch300 denote geometry-predicted arrivals for F4-S01–F4-S06. (b) Channel-base current and relatively calibrated optical-brightness records; colored triangles mark the six retained source times. (c) Colocated infrasound, DAS, and Node-Z waveforms at sites 203 and 204. Infrasound was filtered at 5–45 Hz, whereas DAS and Node-Z records were filtered at 5–120 Hz; no trace-wise normalization was applied. Triangles above (c) mark site-203 acoustic arrivals predicted from source time and three-dimensional source–receiver distance using 343 m $s^{-1}$, without a fitted lag. Time is relative to 08:38:26.5 UTC. Node-Z velocity is displayed in μm $s^{-1}$, whereas DAS strain rate is displayed in nε $s^{-1}$.

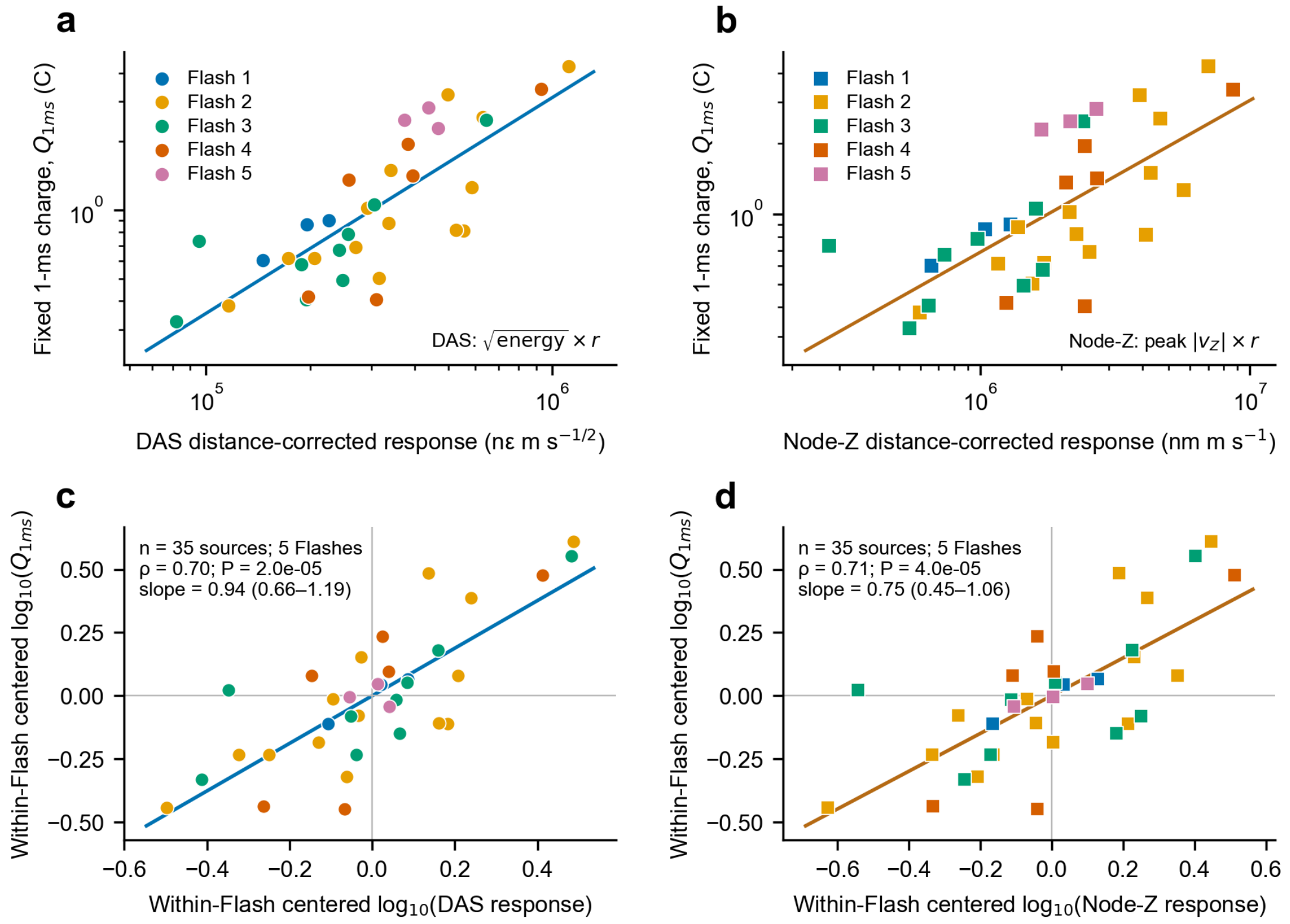

**Figure 3. First-millisecond charge transfer and ground-response associations.** (a) First-millisecond charge transfer, $Q_{1\text{ms}}$, versus the distance-corrected DAS response for 35 sources with analysis-valid, unsaturated current records from five flashes. The analyzed subset comprises 3, 14, 9, 6, and 3 sources from Flashes 1–5, respectively; F3-S05 and F7-S01 are not included. The DAS response was calculated from 5–500 Hz strain-rate records. (b) $Q_{1\text{ms}}$ versus the distance-corrected Node-Z response for the same sources, calculated from response-removed 5–120 Hz vertical-velocity records. Lines in (a) and (b) are descriptive fits to the raw $\log_{10}$-transformed values. (c) Within-Flash relationship between $Q_{1\text{ms}}$ and the DAS response after mean-centering the $\log_{10}$-transformed variables within each flash. (d) Corresponding within-flash relationship between $Q_{1\text{ms}}$ and the Node-Z response using the same procedure. Lines in (c) and (d) show the common within-flash slopes, with parentheses indicating stratified-bootstrap 95% intervals. Spearman rank correlations ($\rho$) were evaluated from the within-flash relationships, with two-sided $P$ values obtained from within-flash permutations.

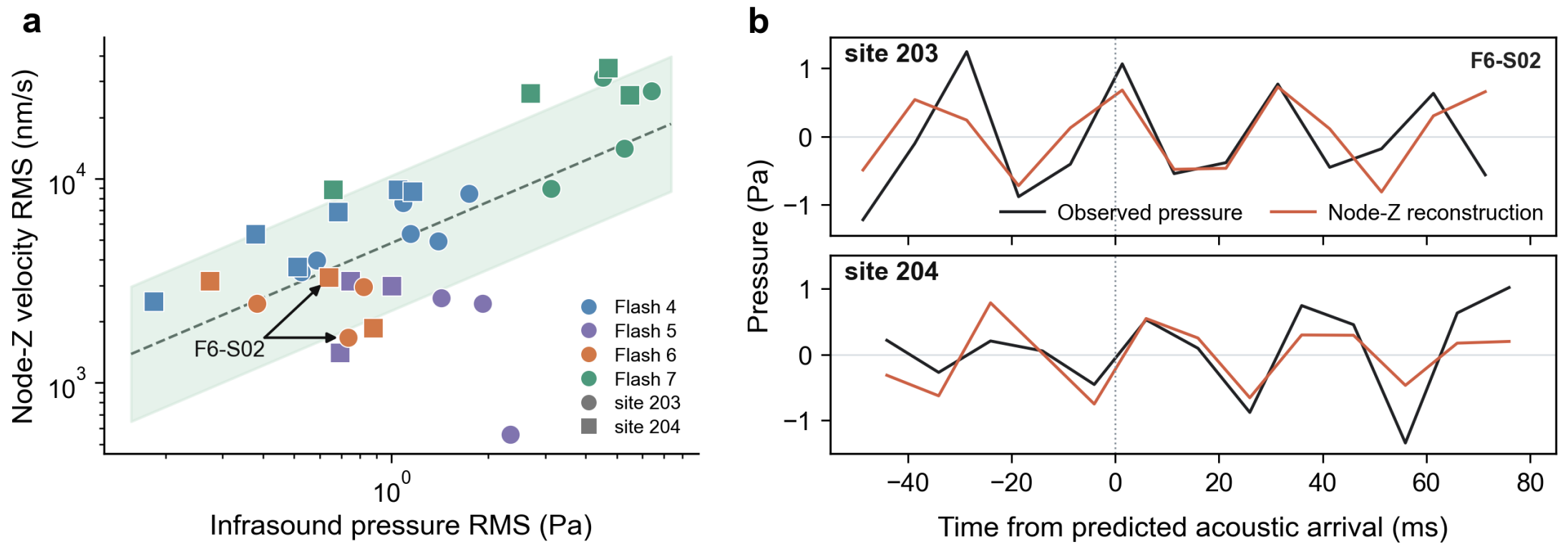

**Figure 4. Narrow-band acoustic–ground coupling and station-specific pressure reconstruction.** (a) Infrasound pressure RMS versus Node-Z vertical-velocity RMS in the common 28–34 Hz band for 32 source–station pairs from 16 temporally resolvable sources at sites 203 and 204. Circles and squares denote sites 203 and 204, respectively, and colors denote Flash 4–7. Arrows identify the two F6-S02 observations whose waveforms are shown in (b). The dashed line is the pooled empirical log–log fit; shading denotes ±1 residual standard deviation in log space. (b) Observed infrasound pressure and Node-Z-reconstructed equivalent pressure in the 28–34 Hz band at sites 203 and 204 for the held-out F6-S02 test. Station-specific complex transfer functions were estimated from Flashes 4, 5, and 7 and applied to Flash 6 without refitting. Traces are shown in pascals without RMS normalization or an additional fitted time shift; time is relative to the geometry-predicted acoustic arrival.

# Supplementary Information for

# Distributed Acoustic Sensing Links Millisecond Lightning-Pulse Charge Transfer to Thunder-Induced Ground Motion

Heting Hong[1*], Yashun Tian[2], Gaopeng Lu[2*], Gan Yang[3], Shaodong Chen[3], Yanfeng Fan[4], Weitao Lyu[4], Jingxiao Liu[1*]

1. School of Civil and Environmental Engineering, Nanyang Technological University, Singapore 639798, Singapore.

2. School of Earth and Space Sciences, University of Science and Technology of China, Hefei 230026, China.

3. Institute of Tropical and Marine Meteorology, China Meteorological Administration, Guangzhou 510640, China,

4. State Key Laboratory of Severe Weather Meteorological Science and Technology / CMA Key Laboratory of Lightning, Chinese Academy of Meteorological Sciences, Beijing, China

**Corresponding authors:**
Dr. Heting Hong (heting.hong@ntu.edu.sg; hardyhong1314@gmail.com);
Prof. Gaopeng Lu (gaopenglu@gmail.com);
Prof. Jingxiao Liu (jingxiao.liu@ntu.edu.sg).

## Supporting Information

**Contents of this file**

- Text S1. DAS location of effective acoustic sources and statistical location uncertainty.
- Figures S1–S5.
- Table S1. Summary of multisensor parameters for the 47 retained sources.

**Introduction**

This file provides additional source-location details, diagnostic figures, and a summary table for the retained thunder sources. Table S1 summarizes source-level quantities; the study data are archived in the Figshare repository cited in the main manuscript.

**Text S1. DAS location of effective acoustic sources and statistical location uncertainty**

The DAS data were sampled at 5 kHz. Before location, each channel was centered by subtracting its temporal median, and the spatial median was removed separately within each contiguous retained cable segment at every time sample to suppress common-mode energy. Channels 0–25, 149–238, and channels above 490 were not used, leaving 375 receivers (Ch26–148 and Ch239–490). Receiver positions were taken from the measured three-dimensional cable geometry. A fourth-order zero-phase Butterworth band-pass filter of 40–450 Hz was used for all seven flashes before calculation of the analytic-signal amplitude. For a trial effective acoustic source at position $r_s$ and origin time $t_0$, the predicted arrival at retained DAS channel i was

$$\hat{t}_i(\mathbf{r}_s, t_0) = t_0 + \frac{\| \mathbf{r}_i - \mathbf{r}_s \|}{c}, \qquad i = 1, \ldots, N,$$

where $r_i$ is the measured three-dimensional channel position and the acoustic speed was fixed at $c = 343$ m $s^{-1}$. The predicted arrivals are geometry-based and not independently hand-picked for each channel. Searches were restricted to event-specific channel–time intervals containing visually coherent acoustic moveout, with additional data read outside some windows only to reduce filter-edge effects.

Source location used a full-array envelope-stack objective that measures how consistently a trial travel-time curve intersects enhanced Hilbert-envelope energy across the usable array. The envelope was robustly normalized by channel and smoothed over a short time window before location at the native 5 kHz sampling rate. For each trial source, one processed feature value was sampled from each valid receiver at its predicted arrival, with isolated large single-channel contributions capped to limit their influence. The basic score was

$$S(\mathbf{r}_s, t_0) = \frac{1}{N_{\text{valid}}} \sum_{i \in \text{valid}} \min\,[F_i(\hat{t}_i), 6],$$

where $F_{i(t)}$ is the processed nonnegative envelope feature, and $N_{valid}$ is the number of receivers contributing to the trial. Fine searches were constrained around an initial coherent moveout so that candidates departing strongly from the seed curve were rejected. The final 47-source catalog was obtained using coarse-to-fine backprojection based on the predicted full-array moveout. When several sources occurred within the same event window, the strongest moveout was suppressed before searching for the next candidate to reduce duplicate detections. For source events with less distinct DAS signatures, high-signal-to-noise arrival anchors were introduced as additional timing constraints.

A candidate was retained as a DAS-located acoustic source only when its predicted arrival curve followed visible coherent moveout across a substantial portion of both usable cable segments, did not simply reproduce a previously extracted strong curve, and did not terminate at a horizontal search boundary. Aligned-stack diagnostics and independent-current timing checks were used for quality control. Geometry-predicted arrivals used throughout the manuscript were calculated directly as source time plus three-dimensional source–receiver distance divided by 343 m $s^{-1}$.

Internal location stability for the full catalog was evaluated with a spatial-block bootstrap using a standardized 40–450 Hz, 5 kHz Hilbert-envelope feature for all seven flashes. The accepted source time, sound speed, cable geometry, and preprocessing choices were held fixed. The 375 retained channels were divided into 10 contiguous blocks; for each replicate, 10 blocks were sampled with replacement, and the source was relocated by a local coordinate-grid search around the accepted solution. A full-channel local optimum served as the reference for the bootstrap offsets. One hundred replicates were generated for each of the 47 sources, yielding 4,700 relocations with equal source weighting. The common horizontal 95% statistical uncertainty radius was defined as the empirical 95th percentile of the pooled horizontal offsets, and its sampling uncertainty was evaluated with 2,000 source-cluster bootstrap iterations. The pooled horizontal RMS offset was 1.205 m, and the common horizontal 95% radius was 2.730 m, with a source-cluster-bootstrap 95% interval of 2.434–2.915 m; the vertical standard deviation was 1.540 m, and the three-dimensional 95% radius was 4.086 m. We therefore report a common horizontal statistical location uncertainty of 2.7

m. This value quantifies internal sensitivity to DAS-channel sampling under fixed model assumptions and is not an absolute location error; it does not include systematic uncertainty from atmospheric sound-speed structure, cable-coordinate errors, source-time uncertainty, non-point-source radiation, or other model inadequacy. The 0.1 m fine-grid spacing used in selected searches is a numerical increment and should not be interpreted as location accuracy.

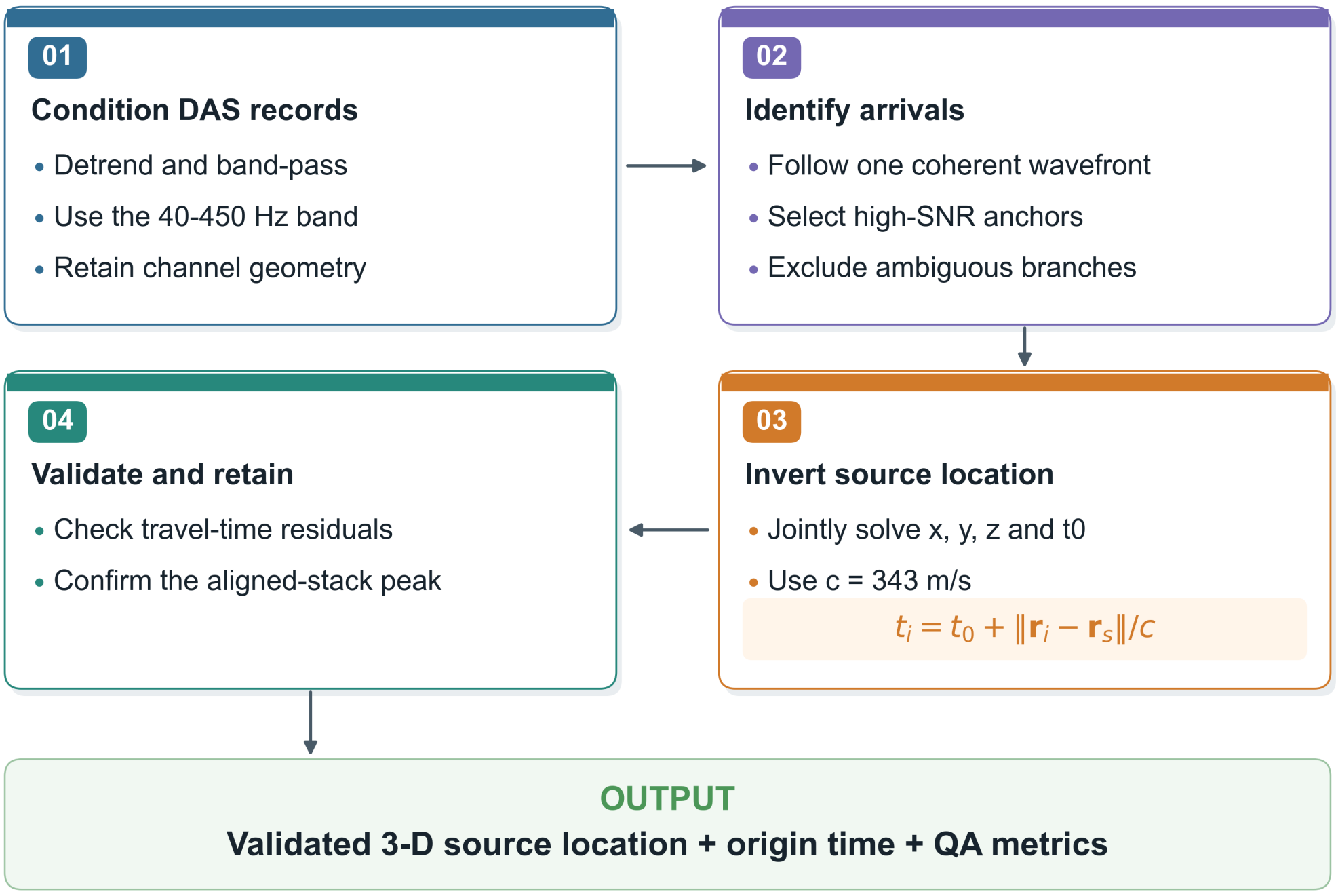

**Figure S1. DAS-based location and validation of effective acoustic sources.** DAS records were temporally median-centered, corrected for segmentwise common-mode energy, band-pass filtered at 40–450 Hz, and transformed to a robustly normalized Hilbert-envelope feature. Geometry-predicted arrivals were calculated from measured three-dimensional cable coordinates using an acoustic speed of 343 m $s^{-1}$. Candidate locations were evaluated against coherent full-array moveout, with additional timing constraints used where applicable. Spatial-block bootstrap relocation yielded a common 95% horizontal statistical location uncertainty radius of 2.7 m for the 47-source catalog; this represents internal sensitivity to DAS-channel selection rather than absolute location accuracy.

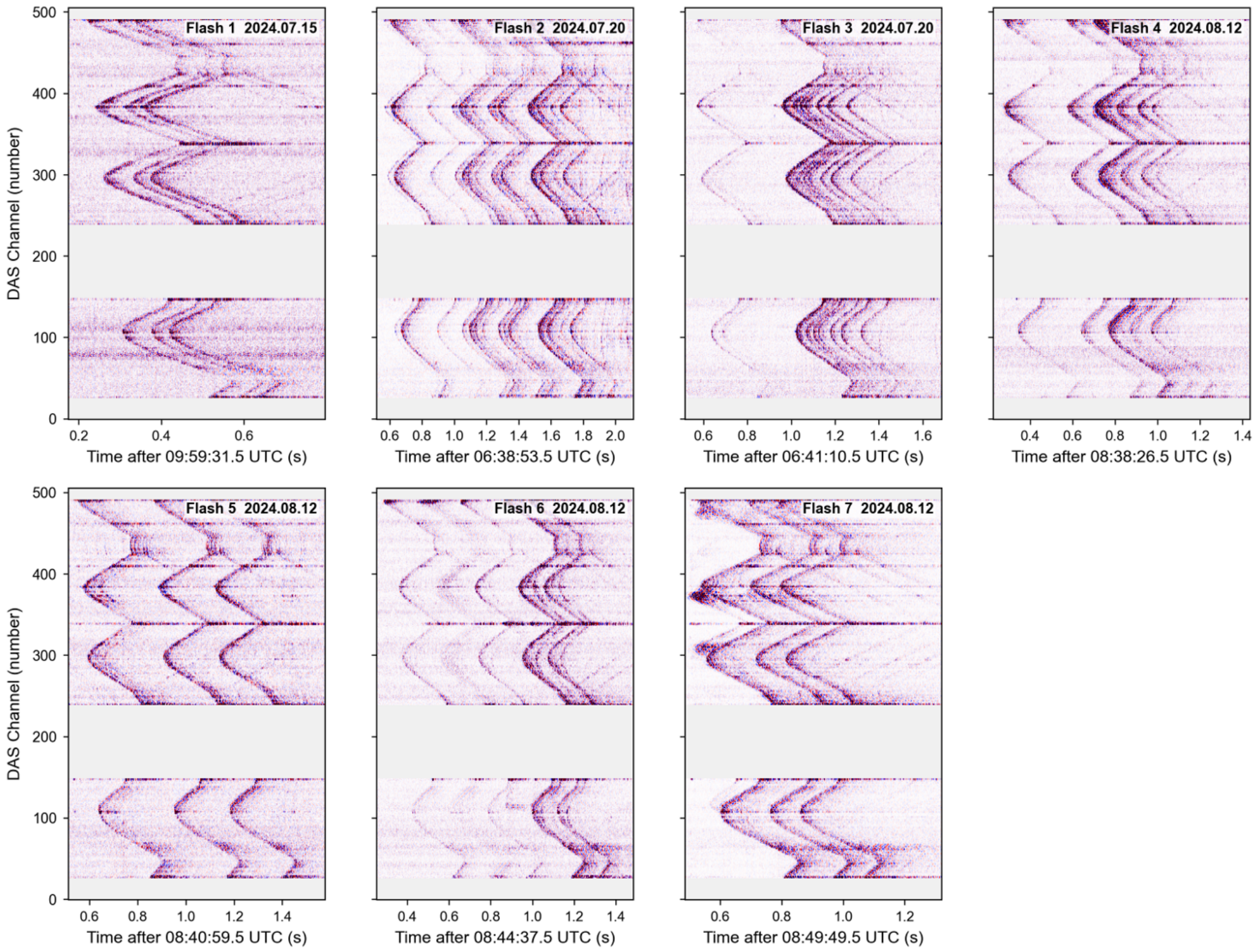

**Figure S2. DAS records for Flashes 1–7.** Signed DAS strain-rate records band-pass filtered at 40–450 Hz show the coherent thunder wavefronts associated with the retained source catalog. Flashes 1–4 are shown in the first row and Flashes 5–7 in the second row. Each panel uses an event-specific UTC reference and display window and is labeled with the flash number and date. Light-gray horizontal gaps mark DAS channels not used for source location.

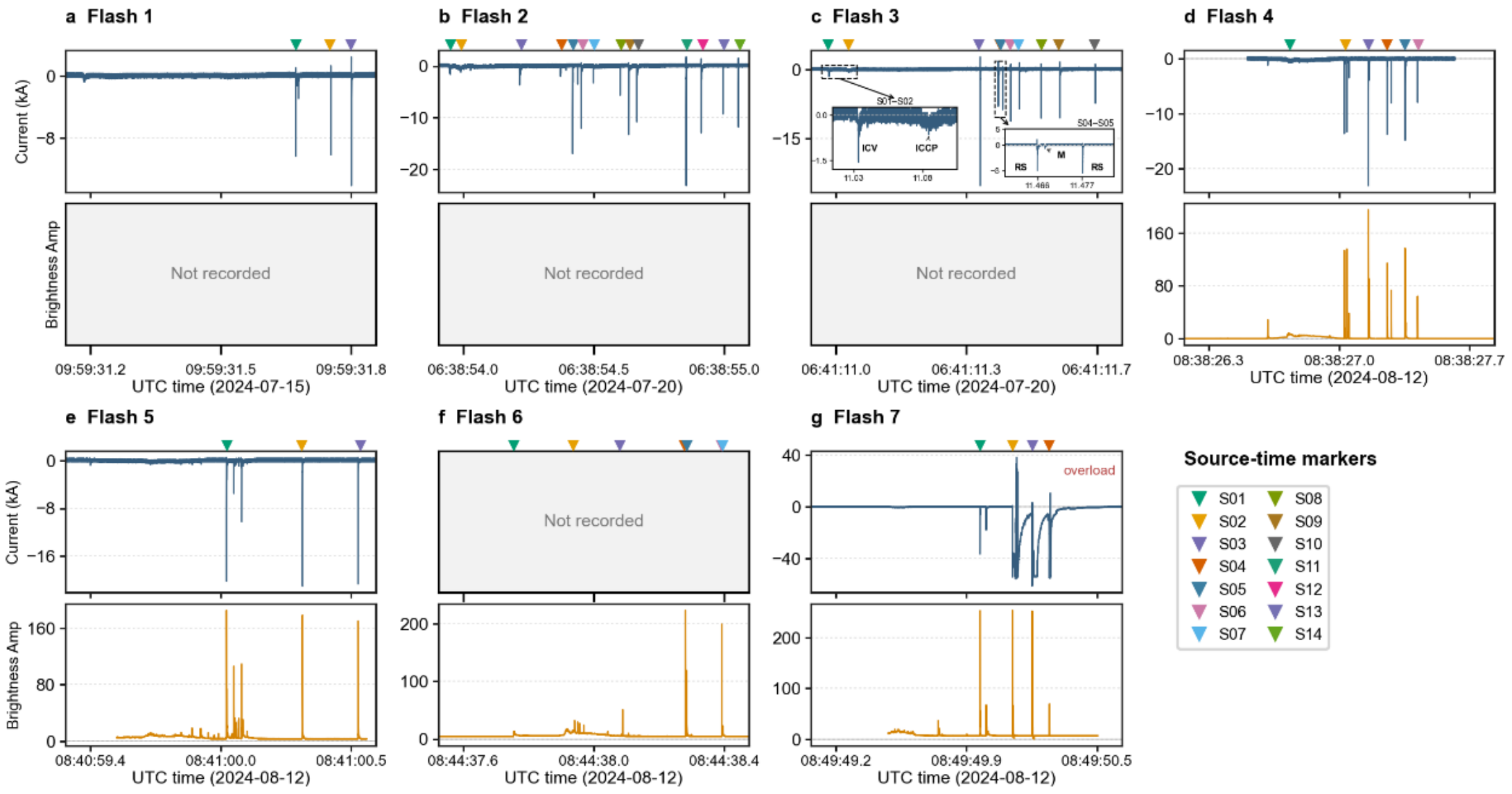

**Figure S3. Current and optical-brightness waveforms for Flashes 1–7.** Each flash panel shows the available channel-base current record above the optical-brightness record, with both plotted against UTC time. Colored inverted triangles identify the retained source times. The two insets in Flash 3 enlarge the early ICV and ICCP events (S01–S02) and the interval containing S04, S05, and the intervening weak pulse. Dashed rectangles indicate the enlarged intervals, and connecting arrows link them to the corresponding insets. Labels identify initial current variation (ICV), initial continuous current pulses (ICCPs), return strokes (RSs), and an M-component (M). Gray panels marked "Not recorded" indicate missing measurements and should not be interpreted as zero signal. Optical brightness is reported in relative amplitude units, and overloaded portions of the Flash 7 current record are not quantitatively valid.

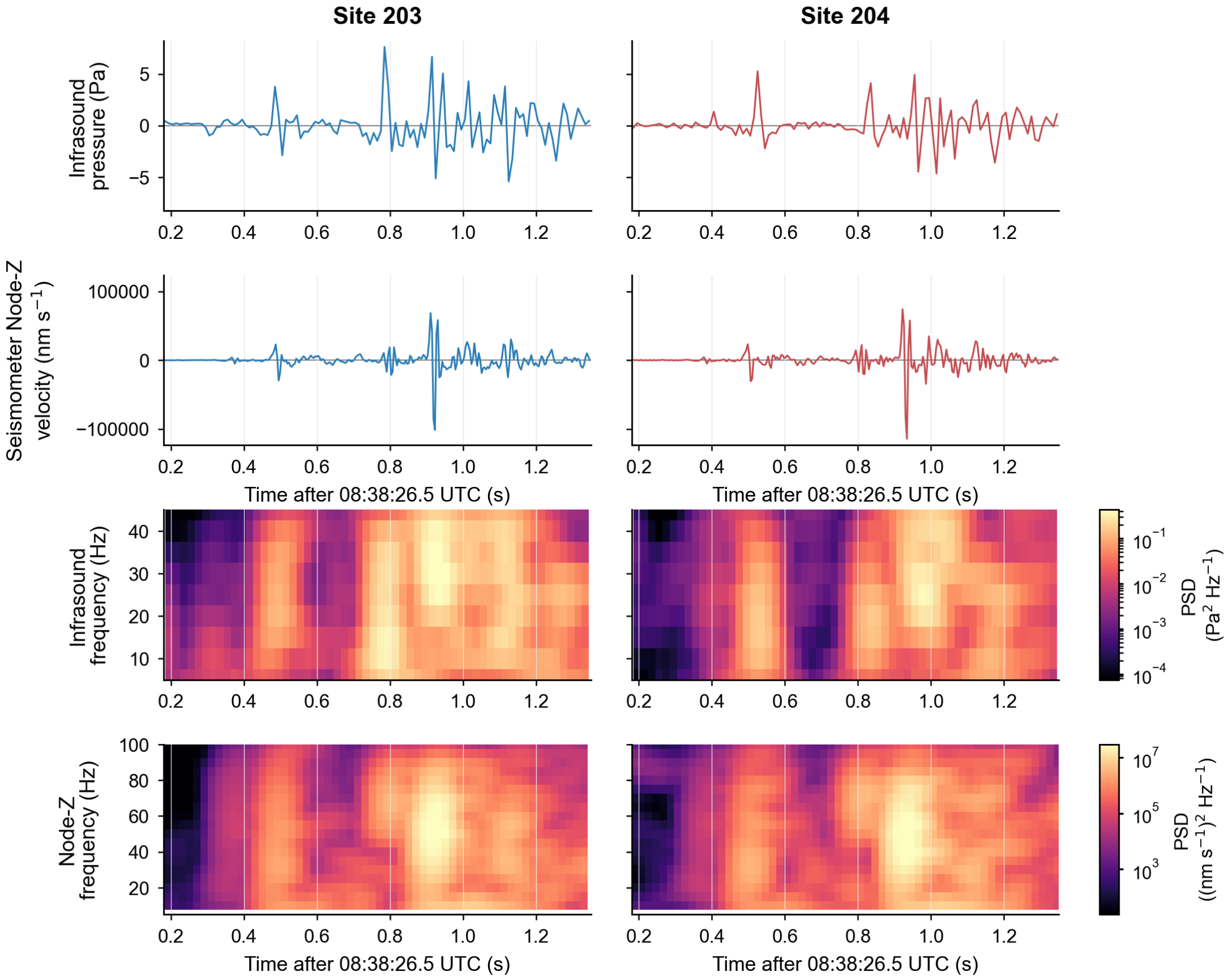

**Figure S4. Flash 4 infrasound and Node-Z waveforms and spectrograms.** At sites 203 and 204, pressure and SAC-native vertical velocity are shown for F4-S01–F4-S06 over 08:38:26.680–08:38:27.850 UTC (Figure 2c). Records retain native sampling (100 Hz and 250.0 Hz), are detrended and band-pass filtered at 5–45 Hz (infrasound) and 5–100 Hz (Node-Z), and are not trace-normalized. PSDs use 0.20 s Hann windows (90% overlap; 5 Hz bins); displayed fields are lightly Gaussian-smoothed. Color limits are shared between sites within each sensor type. Waveform units are Pa and nm $s^{-1}$; PSD units are $Pa^2$ $Hz^{-1}$ and (nm $s^{-1}$) $^2$ $Hz^{-1}$.

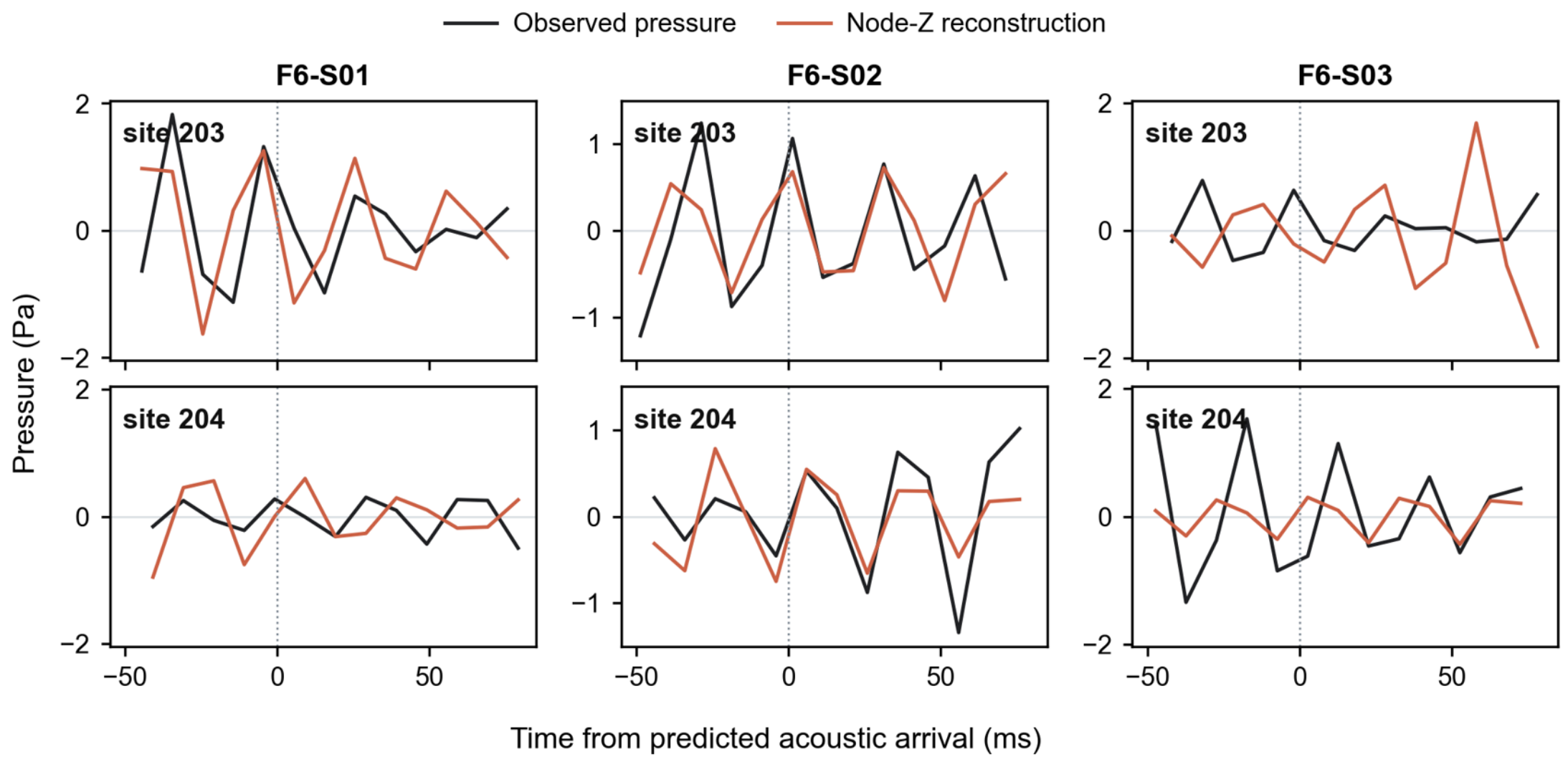

**Figure S5. Station-specific pressure reconstruction for individually resolvable Flash 6 sources.** Observed 28–34 Hz infrasound pressure (black) and Node-Z-reconstructed equivalent pressure (orange) are shown at sites 203 and 204 for F6-S01, F6-S02, and F6-S03. Station-specific complex transfer functions were estimated from Flashes 4, 5, and 7 and applied to the held-out Flash 6 without refitting. Vertical dotted lines mark the geometry-predicted acoustic arrivals. Traces are shown in pascals without RMS normalization or an additional fitted time shift. F6-S04–S07 are not shown individually because the 100 Hz pressure records do not resolve the S04/S05 and S06/S07 arrival pairs as separate pressure arrivals.

**Table S1. Summary of source-level multisensor parameters for 47 DAS-located thunder source events in seven flashes.**

| Flash | Source | DAS √Energy × r (×10⁵ nε m / √s) | Node-Z peak \|vZ\| × r (×10⁶ nm m s⁻¹) | Current Q1ms (C) | Peak current (kA) | Brightness peak (Normalized Amp) |
|---|---|---|---|---|---|---|
| **Flash 1** | S01 | 1.96 | 1.04 | 0.864 | 10.4 | *NaN* |
| | S02 | 1.46 | 0.658 | 0.603 | 10.2 | *NaN* |
| | S03 | 2.27 | 1.3 | 0.906 | 14.2 | *NaN* |
| **Flash 2** | S01 | 1.16 | 0.596 | 0.38 | 1.7 | *NaN* |
| | S02 | 5.55 | 4.12 | 0.815 | 1.13 | *NaN* |
| | S03 | 1.73 | 1.16 | 0.614 | 3.77 | *NaN* |
| | S04 | 2.05 | 1.73 | 0.616 | 2 | *NaN* |
| | S05 | 4.98 | 3.89 | 3.21 | 17.1 | *NaN* |
| | S06 | 3.41 | 4.28 | 1.5 | 12.2 | *NaN* |
| | S07 | 2.7 | 2.54 | 0.689 | 3.27 | *NaN* |
| | S08 | 3.37 | 1.38 | 0.879 | 5.7 | *NaN* |
| | S09 | 2.93 | 2.14 | 1.02 | 13.4 | *NaN* |
| | S10 | 5.86 | 5.66 | 1.26 | 10.7 | *NaN* |
| | S11 | 11.2 | 7.03 | 4.28 | 23.3 | *NaN* |
| | S12 | 6.31 | 4.65 | 2.57 | 13 | *NaN* |
| | S13 | 3.16 | 1.56 | 0.503 | 9.37 | *NaN* |
| | S14 | 5.28 | 2.28 | 0.819 | 11.9 | *NaN* |
| **Flash 3** | S01 | 0.955 | 0.274 | 0.733 | 1.63 | *NaN* |
| | S02 | 0.82 | 0.545 | 0.324 | 0.667 | *NaN* |
| | S03 | 6.45 | 2.42 | 2.49 | 25.3 | *NaN* |
| | S04 | 1.89 | *1.7* | 0.577 | 8.1 | *NaN* |
| | S05 | 1.72 | *1.18* | 0.501 | 8.83 | *NaN* |
| | S06 | 2.58 | 0.978 | 0.784 | 11.4 | *NaN* |
| | S07 | 2.47 | 1.45 | 0.495 | 8.7 | *NaN* |

| Flash | Source | DAS √Energy × r (×10⁵ nε m / √s) | Node-Z peak \|vZ\| × r (×10⁶ nm m s⁻¹) | Current Q1ms (C) | Peak current (kA) | Brightness peak (Normalized Amp) |
|---|---|---|---|---|---|---|
| | S08 | 2.42 | 0.737 | 0.67 | 10.7 | *NaN* |
| | S09 | 3.06 | 1.61 | 1.06 | 10.6 | *NaN* |
| | S10 | 1.94 | 0.645 | 0.406 | 7.43 | *NaN* |
| **Flash 4** | S01 | 3.1 | 2.44 | 0.405 | 0.667 | 8.66 |
| | S02 | 2.58 | 2.08 | 1.36 | 13.7 | 136 |
| | S03 | 9.31 | 8.69 | 3.41 | 23.2 | 195 |
| | S04 | 3.96 | 2.72 | 1.42 | 13.8 | 114 |
| | S05 | 3.82 | 2.44 | 1.95 | 15 | 137 |
| | S06 | 1.97 | 1.25 | 0.417 | 7.97 | 64.3 |
| **Flash 5** | S01 | 3.75 | 2.16 | 2.5 | 20.3 | 185 |
| | S02 | 4.39 | 2.69 | 2.82 | 21.1 | 178 |
| | S03 | 4.69 | 1.68 | 2.29 | 20.7 | 170 |
| **Flash 6** | S01 | 1.17 | 1.84 | *NaN* | *NaN* | 13.3 |
| | S02 | 0.823 | 1.29 | *NaN* | *NaN* | 32.2 |
| | S03 | 1.27 | 1.41 | *NaN* | *NaN* | 51.3 |
| | S04 | 3.68 | 3.84 | *NaN* | *NaN* | 224 |
| | S05 | 1.96 | 7.55 | *NaN* | *NaN* | 119 |
| | S06 | 1.69 | *2.15* | *NaN* | *NaN* | 200 |
| | S07 | 0.913 | *4.43* | *NaN* | *NaN* | 22.8 |
| **Flash 7** | S01 | 12.4 | 12.5 | 16.1 | 36.8 | 253 |
| | S02 | 11.1 | 8.35 | *NaN* | *NaN* | 254 |
| | S03 | 10.8 | 8.91 | *NaN* | *NaN* | 252 |
| | S04 | 4.44 | 3.3 | *NaN* | *NaN* | 69.5 |

**Note.** Units are $10^5$ nε m / √s (DAS), $10^6$ nm m $s^{-1}$ (Node-Z), C (charge), kA (peak current), and relative amplitude (brightness). NaN denotes unavailable or invalid data. Figure 3 uses Flashes 1–5, excluding F3-S05; F7-S01 is outside this subset.